Going Viral: An Analysis of Advertising of Technology Products on TikTok

December 11, 2023

Ekansh Agrawal

Department of Interdisciplinary Studies Field

University of California, Berkeley

Advisor: Dr. Fang Xu

Second Reader: Dr. Shreeharsh Kelkar



## Table of Contents





**ABSTRACT**

Social media has transformed the advertising landscape, becoming an essential tool for reaching and connecting with consumers. Its sharing and engagement features amplify brand exposure, while its cost-effective options provide businesses with flexible advertising solutions. TikTok is a more recent social media platform that has gained popularity for advertising, particularly in the realm of e-commerce, due to its large user base and viral nature. TikTok had 1.2 billion monthly active users in Q4 2021, generating an estimated $4.6 billion revenue in 2021. Virality can lead to a massive increase in brand exposure, reaching a vast audience that may not have been accessible through traditional marketing efforts alone. Advertisements for technological products are an example of such viral ads that are abundant on TikTok. The goal of this thesis is to understand how creators, community activity, and the recommendation algorithm influence the virality of advertisements for technology products on TikTok. The study analyzes various aspects of virality, including sentiment analysis, content characteristics, and the role of influencers. It employs data scraping and natural language processing tools to analyze metadata from 2,000 TikTok posts and 274,651, offering insights into the nuances of viral tech product advertising on TikTok.

**INTRODUCTION**

Viral marketing is a technique that leverages networks of users to spread information exponentially. With the advent of social media, the potential of viral marketing has expanded massively. Platforms such as Twitter, Instagram, Facebook, and TikTok have become vital arenas for viral content dissemination. These platforms, with their billions of users, create an



environment where content can be shared, liked, and retweeted at an exponential rate. The very nature of social media, where content consumption is a daily routine for many, provides fertile ground for marketers to introduce campaigns that can easily resonate with audiences. Moreover, the interactive element of social media means that consumers can not only consume content but also participate in its spread, adding comments, twists, and personalized touches. This two-way engagement ensures a deeper connection between the brand and its audience.

All major social media platforms have their own advertisement pipelines that allow influencers and businesses to sponsor content in order to increase their reach. The goal of this paper is to delve into the viral advertising of technology products on TikTok. The emphasis on technology products stems from the surge of innovative gadgets showcased on social media, many of which, despite their unique features, often have counterparts with similar functionalities. TikTok's concise content format offers an advantage, allowing for brief yet impactful product reviews that cater to contemporary attention spans more effectively than extended video analyses. TikTok's novelty in the social media space, as well as its exponential user growth (Chen), furnishes a ground to cultivate a rich dataset. The novelty of the platform enables exploration into the specific characteristics that propel technology-related content to viral status on TikTok.

TikTok is designed to encourage user engagement. Viewers can like, comment, share, and follow creators. Engagement metrics, like the number of likes, comments, and shares, are prominently displayed on each video. The "For You Page" (FYP) is TikTok's main feed, where users discover new content tailored to their interests based on the algorithm's recommendations. The goal for creators is to have their content featured on the FYP for increased visibility. The



recommendation algorithm is core to every social media platform; it's responsible for providing users with fresh content on their page while maximizing user engagement over time. TikTok's algorithm stands out for several reasons. It emphasizes short video clips and aims to boost emerging and lesser-known creators, fostering a broader variety of content on the platform (Zhang & Liu, 2021). The algorithm takes into consideration videos users have interacted with, accounts and hashtags they follow, their location and language preferences, and the type of content they create (Worb 2023). Moreover, the algorithm is adept at pinpointing popular content based on current viral audio, hashtags, and comments. Having prior viral content is not a factor the algorithm considers, but it's evident that such content has features that are looked on favorably by the recommendation algorithm. By showcasing content to a larger audience, the TikTok algorithm has been leading to the swift proliferation of viral trends and challenges on the platform (Zhang & Liu, 2021).

Advertisements on TikTok can often blend seamlessly into the regular content flow. This is because many of these ads are created by influencers and content creators who style their promotional content in a way that mimics their usual posts. This approach makes the ads more engaging and less intrusive, as they appear as a natural part of the user's scrolling experience. Concentrating on these types of posts is crucial for this research because it uncovers the subtle details that contribute to a post's virality. More often than not, posts with an overt *#ad* tags tend to have comments disabled, likely as a strategy to avoid potential controversy. Focusing on the less obtrusive products spots or advertisements will allow this research to highlight the different engagement strategies between openly advertised content and more subtly promoted material, providing valuable insights into the dynamics of online interactions and marketing effectiveness.



**BACKGROUND**

<u>Viral Marketing</u>

Kaplan and Haenlein (2011) define viral marketing as a strategy that entices users to willingly or unwillingly spread marketing messages to others. The widespread adoption of social media platforms like Facebook, Twitter, and YouTube has made them essential mediums for viral content dissemination.

Viral advertisements on TikTok present a fascinating duality in terms of their benefits and challenges. On one hand, the benefits are hard to ignore. They offer a cost-effective avenue for brands to promote themselves, primarily due to the low expenditure associated with such campaigns. Additionally, the platform's vast user base ensures a wide reach, thereby increasing the potential audience size for any given advertisement. Once an advertisement gains traction, its dissemination can be incredibly rapid, often spreading across various demographics and regions in a short period. This speed, coupled with the interactive nature of the platform, fosters deep customer engagement, as identified by Phelps et al (2004). However, the very factors that make TikTok a promising medium for viral marketing also contribute to its challenges. There exists a significant loss of control over the message once it goes viral, making the outcomes quite unpredictable. As Stephen points out, this unpredictability can sometimes lead to negative virality, where the brand might receive unintended backlash from audiences. Furthermore, the ephemeral nature of trends on platforms like TikTok means that campaigns can be short-lived, making sustained engagement a challenge for marketers.



Berger in his book *Contagious: Why Things Catch On*, pinpoints six key factors that make content more likely to go viral: social currency, triggers, emotion, public, practical value, and stories. The emotional intensity of the content, be it positive or negative, plays a critical role in virality. Articles that evoke high-arousal emotions such as awe, anger, or amusement are more likely to be shared (Berger and Milkman). Content that evokes visceral emotional reactions is more likely to be shared with people who share commonalities in values and belief systems (Berger). Content that creates a conjoined feeling, whether it be validation or disgust, is more likely to be shared within and across circles.

Virality is often measured using metrics such as share counts, view counts, and engagement rates. Goel argued that true virality is rare and that most content labeled "viral" achieves its reach through a broadcast pattern rather than a purely person-to-person spread. This sentiment was echoed by a study that aimed to improve the detection of viral tweets, especially focusing on those containing potentially misleading content (Elmas). Researchers identified a potential metric for virality: the ratio of retweets to the author's followers. A tweet is more likely to be deemed viral if the ratio of its retweets to the author's followers surpasses a threshold of 2.16 (Elmas).

Expansive social media marketing strategies are the core of how firms create strategic objectives by analyzing resources, capabilities, and market structure. The main strategies employed are commerce, content, and monitoring (Li). A social commerce strategy focuses on transactional activities on social media, primarily aiming to sell. Social content strategy centers on creating and distributing valuable content to attract or retain customers. A social monitoring strategy involves listening and responding to customer interactions on social media, aiming to



enhance customer satisfaction. The very nature of their interaction directly influences the desired level of customer engagement.

<u>Word of Mouth</u>

Electronic Word of Mouth (eWOM) refers to the process of consumers sharing information, opinions, and recommendations about products, services, brands, or companies through electronic channels, primarily on the Internet and social media platforms. It is essentially the digital version of traditional word-of-mouth marketing, where people discuss their experiences and preferences, but it occurs in online spaces and often has a broader reach and faster dissemination (Chevalier). Studying eWOM can provide us insight in terms of studying virality and understanding how content disseminates over social media. With the comments section in TikTok, it's important to consider factors such as popular phrases that might affect the way eWOM helps or hinders the information flow.

A study by Phelps sought to understand the complex interplays behind pass-along emails, an email that is forwarded or shared by the recipient with others, typically because they find the content interesting, informative, or relevant. The study categorizes different types of emails that get passed along, such as jokes related to gender issues, work, or computers; warnings about crimes; games; chain letters related to luck or religion; and messages offering free products or services (Phelps). The study notes that people may share content to provide information to others, to enhance their social status by appearing knowledgeable or helpful, or to receive incentives such as discounts or rewards. The paper also suggests that the reactions to receiving pass-along emails are interdependent. For instance, the decision to open or delete an email can



influence reactions at other stages. If an email is deleted, the episode ends. The study also delved

into the role of personalization in pass-along emailing, suggesting that personal touches might

influence the decision to forward (Phelps). Even though community activity on TikTok is about

engaging with a broader audience in a more interactive, communal space (de La Rocha), I still

think I can draw upon direct communication to analyze how to foster engagement from viewers..

Consumer Intentions

      In the realm of e-commerce, successful advertisements can convince their target audience

that the advertised service/good is a worthy purchase. In a study aimed to analyze the critical

significance of social media in daily routines, including its impact on consumers' purchase

intention in Jakarta, Indonesia (Permatasari), researchers contextualized advertising on social

media concerning the 3 main factors which are perceived price, perceived risk, and perceived

value. The study states that specific elements play pivotal roles in a consumer's decision to buy

or refrain from purchasing based on an advertisement. "Perceived price" denotes the cost

consumers deem fitting for a product, influenced by factors like product quality, brand image,

and the product's perceived advantages. Conversely, "perceived risk" signifies the uncertainty or

potential setbacks consumers sense when considering a purchase, encompassing worries about

product quality, seller trustworthiness, and transaction security. Lastly, "perceived value" is the

perceived benefits associated with a product relative to its price. The researchers focused on

repeat customers of e-commerce sites and examined the mediating role of perceived risk and

perceived value in the relationship between social media and purchase intention. The study found

that the highest impact on consumers' purchase intention was found to be perceived value



(Permatasari). What the audience assumes to be the value of the product being advertised is a crucial driver of online transactions in e-commerce businesses. By looking at comments on TikTok viewers can understand what other people are saying about the product, which can directly affect whether or not they want to buy the product. While it's difficult to ascertain exactly whether or not a user bought a product they saw, I can strive to understand buyer confidence with a product or service.

Cultural content on social media refers to the collective expressions, symbols, practices, and narratives that represent the shared values, beliefs, and customs of specific groups or societies. This content can manifest in various forms, such as memes, videos, stories, or images, reflecting the intricate weave of cultural identities and communal understandings. In the realm of social media advertising, understanding and leveraging cultural content is crucial for virality. Advertisements that resonate with cultural norms or evoke familiar cultural narratives have a higher likelihood of being shared, discussed, and engaged with. When advertisers tap into these shared symbols and meanings, they not only increase their reach but also establish a deeper connection with their target audience, enhancing the efficacy of their message (Jenkins).

There is a prevalence of products and services failing if they don't reflect the values of their target audience. With a growing demand for sustainable advertising, there seems to be a gap concerning social media and cross-cultural studies. A study by Minton Et Al (2012) involved 1,018 participants who completed an online survey hailing from the United States, Germany, and South Korea. The study found different motives based on different countries like the United States valuing anti-materialistic behavior while Germany cared more about recycling. Social



media can be used to provide consumers with information about sustainable products and foster a sense of community around sustainable behaviors.

Researchers performed a  quantitative content analysis of 445 viral video ads shared online between 2009 and 2019 (Segev). They identified several key factors influencing the virality of video ads. Emotionally charged videos, especially those that elicit strong positive emotions like humor or cuteness, are more likely to be shared, as are those that provoke high arousal emotions, regardless of whether they are positive or negative. The structure of the content also matters; videos with a well-developed storyline, encompassing exposition to denouement, have a higher likelihood of being shared (Segev). Creativity is a significant driver of virality, with uniquely creative ads capturing more attention and shares. Furthermore, while information-focused content might sometimes deter sharing, there are exceptions, especially if the content is perceived as risky. An underlying theme is the role of social identity in sharing decisions. Individuals tend to share content that positively reflects their social image, emphasizing the importance of creating content that resonates with the sharer's identity. Ads that foster high consumer engagement due to their interactivity and shareability are more likely to spread virally (Segev) .

A study by Kulkarni et al. studied the potential influence of consumers' personalities and how their variability affects their information processing and their corresponding sharing tendencies (2020). The study categorized people based on their need for cognition as well as advertisements based on whether they were informational or emotional. Upon showing different advertisements to the study's participants, they reported attitudes towards the viral ad, emotional arousal, message complexity, product involvement, and celebrity endorser's credibility. The



results revealed that high-NFC individuals tend to share informational viral ads more than emotional ones, while low-NFC individuals show the opposite pattern (Kulkarni). This challenges the traditional Elaboration Likelihood Model (ELM) framework. ELM is a theory in persuasion communication that suggests there are two primary routes to persuasion: the central pathway and the peripheral pathway. The central pathway involves thoughtful consideration of the arguments presented, requiring more cognitive effort and leading to lasting attitude change. This pathway is taken when the audience is motivated and able to process the information. In contrast, the peripheral pathway relies on superficial cues or heuristics, such as the attractiveness of the speaker or the number of arguments rather than their quality. This pathway is taken when the audience is not motivated or able to think deeply about the message, leading to temporary attitude change (Petty).

Influencer Community

      In recent years, influencers have emerged as powerful figures in the digital landscape. Djafarova and Rushworth (2017) found that influencers, due to their perceived authenticity and relatability, can significantly impact consumer behavior. Their study suggests that influencers can make advertisements more relatable, leading to higher engagement and virality. User-generated content (UGC) has been identified as a key factor in driving virality. Nelson- Field et al. (2013) found that UGC, especially when created by influential creators, can be more engaging and shareable than brand-generated content. This suggests that creators play a pivotal role in the virality of advertisements on social media. A majority of content on TikTok consists of creators remixing existing content and reacting or providing commentary. Putting this twist to it makes



the material more agreeable and organic. Creators often build trust with their audience over time. Lou and Yuan (2019) argue that this trust can be leveraged by brands to create more authentic and believable advertisements. When creators are involved, the perceived authenticity of the advertisement can increase, leading to higher chances of virality.

Creators often have vast networks on social media platforms. De Vries et al. (2012) found that the network size and engagement level of a creator can significantly influence the virality of content. Advertisements shared by creators with large, engaged networks are more likely to go viral. These creators are often skilled storytellers. Berger and Milkman (2012) found that content that evokes strong emotions, whether positive or negative, is more likely to be shared. Creators, with their unique voices and perspectives, can craft advertisements that resonate emotionally with audiences, increasing the likelihood of virality. With TikTok being a platform for short-form videos, brevity is of utmost importance in making digestible content.

TikTok Platform

Different platforms serve different purposes in the viral marketing ecosystem. For example, Zarrella (2009) found that Twitter's "Retweet" function significantly aided virality due to its frictionless nature. Platforms like Facebook and Instagram, which prioritize visual content, can be pivotal for brands that utilize visually appealing or emotional campaigns (Bakhshi, Shamma, and Gilbert). Each social media platform has unique characteristics influencing its role in viral marketing. Facebook, with its vast user base of over 2 billion monthly active users, often serves as a primary news source for 45% of U.S. adults. This extensive reach makes it a critical platform for viral content, particularly for news-related or information-heavy campaigns.



Instagram, surpassing 1 billion users, emphasizes visual content, making it ideal for campaigns that rely on strong visual elements and storytelling. It's especially effective for turning passions into purchases, with 75% of users taking action after viewing a brand's post. While platforms like LinkedIn and Pinterest might not be the first choices for viral success, they offer unique advantages. LinkedIn's user base of over 106 million monthly users, predominantly professionals, makes it suitable for B2B marketing and social selling. Pinterest, with its vast collection of over 100 billion pins and a user base seeking brand content, can also play a vital role in driving website traffic and engagement in specific niches" (Quesenberry).

TikTok videos are usually between 15 seconds and 3 minutes in length, with most content falling within the 15 to 60-second range. TikTok videos are designed to be viewed in a vertical (portrait) orientation, optimized for mobile devices. The recommended aspect ratio is 9:16. Music is a significant part of TikTok content. Users can add music tracks or audio clips to their videos from TikTok's extensive library, and many trends on the platform are tied to specific songs or sound bites. TikTok offers a wide range of video editing tools and effects that users can use to enhance their videos. These include filters, transitions, text overlays, stickers, and more. Similar to other social media platforms, TikTok uses hashtags to categorize and discover content. Creators often include relevant hashtags in their video captions to increase visibility and engagement. Creators can add captions, text, or subtitles to their videos to provide context or convey a message. This text is typically overlaid on the video. TikTok often features challenges and trends, where users create content based on a specific theme or idea. Participating in these trends can help videos gain visibility. TikTok allows users to "duet" with another video, which



means they can create a split-screen video alongside an existing video. There's also a "stitch" feature that lets users incorporate a part of another video into their own.

## METHODOLOGY

TikTok Data

TikTok has offered a Display API since 2018, that allows users to pull public information for a TikTok page based on some query commands. The offerings of this API are quite sparse and provide little to no research functionality in terms of creating and extracting data-rich queries. As of February 2018, TikTok started offering a research API to allow researchers from non-profit universities in the US and Europe to study public data about TikTok content and accounts. Through the research API, researchers should be able to access public data on all accounts and content. However, due to some access limitations and short deadlines for this research project, I was unable to use the official research API for this study. Instead, I opted in to use Apify, which is a cloud platform for web scraping and browser automations, for collecting data. This paid platform consists of third-party agents, which are brute-force automations that aim to reverse-engineer the public API calls that each platform makes for accessing and rendering its resources. Using some predefined agents for TikTok, I was able to define rich queries for scraping and downloading posts, comments, content, and any related metadata.

**Dataset**

|  | **#fyp** | **#tech** | **#techtok** | **#techtips** |
|---|---|---|---|---|
| **Content Characteristics** | General content that aims to attract a wide audience base | General content that consists of technology products or services | New technology products or services, lifestyle blogging | Breaking news pertaining to technology products or services/ expose/ |



| | | | | advice |
|---|---|---|---|---|
| **Audience Type** | Consumers of regular content | General consumers with a mild interest in technology products | People interested in tech products and lifestyle blogging | People interested in purchasing or trying new products |

**Table 1:** *Breakdown of hashtags related to TikTok posts about technology products and some of their intrinsic traits to the platform*

For constructing our dataset, I used a few tools on Apify in order to scrape all the relevant data that I needed to conduct our study. I first used an scraper to download the most 2,000 most popular posts from May 2023 to July 2023 that had at least 3 of the following hashtags: *#fyp*, *#tech*, *#techtok*, or *#techtips*. The restrictive date range for the data was a limitation of the API platform. With these posts I pulled all relevant metadata like creators profiles, number of shares, number of comments, number of plays, and number of likes. I then used another scraper to download the comments for a sample of these posts. Due to budget and time constraints, I was only able to scrape further content out of 1,400 of the original 2,000 scraped posts. Out of these 1,400 posts I attempted to scrape 100-200 comments and up to 10 replies for each comment for a total of 274,651 comments. The number of comments and replies scrapped per post was also a reflection of the budget constraints on Apify. I then used 2 separate agents on the same subset of posts to scrape video/slideshows along with their corresponding audio files and relevant metadata.

In general, the practice of data scraping from social media platforms is considered legal, and TikTok does not actively prevent third-party scrapers from accessing their platform. The scraped data consists solely of publicly available information on TikTok, which any regular TikTok user can access. This scraped data is securely stored on a private encrypted file server.



It's important to note that specific user attributes like political views or nationalities are not directly considered in this analysis. To apply the findings of this paper, all user information has been anonymized, and precautions have been taken to prevent any form of back-tracing.

<u>Natural Language Processing Toolkits</u>

In order to perform some large-scale sentiment analysis on the comments, I opted to use 2 different popular natural language processing toolkits: the *go_emotions* model and NLTK VADER. The *go_emotions* sentiment analysis tool, based on Reddit data, is a language model that utilizes a dataset with 28 distinct emotion labels, making it suitable for capturing nuanced emotional responses in text. It operates as a multi-label classification model, meaning that it can assign one or more emotion labels to a single input text, providing a more comprehensive understanding of sentiment. For practical use, a common approach is to set a threshold of 0.5 on the model's probability outputs, allowing for the prediction of emotions based on the most likely labels given the input text. This versatile tool enhances sentiment analysis by offering a broader range of emotional insights for TikTok comments and other text-based content. The NLTK (Natural Language Toolkit) VADER (Valence Aware Dictionary and sEntiment Reasoner) sentiment tool is a pre-built sentiment analysis tool specifically designed for analyzing the sentiment of text data, primarily in the context of social media and short, informal text. VADER is unique because it is specifically tuned to handle sentiments expressed in social media text, which often includes informal language, slang, and emojis. VADER not only assigns a sentiment polarity (positive, negative, or neutral) to a text but also provides a sentiment intensity score. This score indicates how strong the sentiment is within the text. VADER can handle emojis and



considers capitalization and punctuation for sentiment analysis. For example, an exclamation mark in a sentence can indicate heightened sentiment. For the sake of this research, I accept the sentiment analysis score that VADER predicts if the sentiment intensity score is above a certain threshold. All toolkits work on the English language, and the majority of the comments that were scrapped were also all in English. Comments in any other foreign languages were all translated to English before running any of the natural language processing toolkits.



# DATA ANALYSIS

## Overall Virality

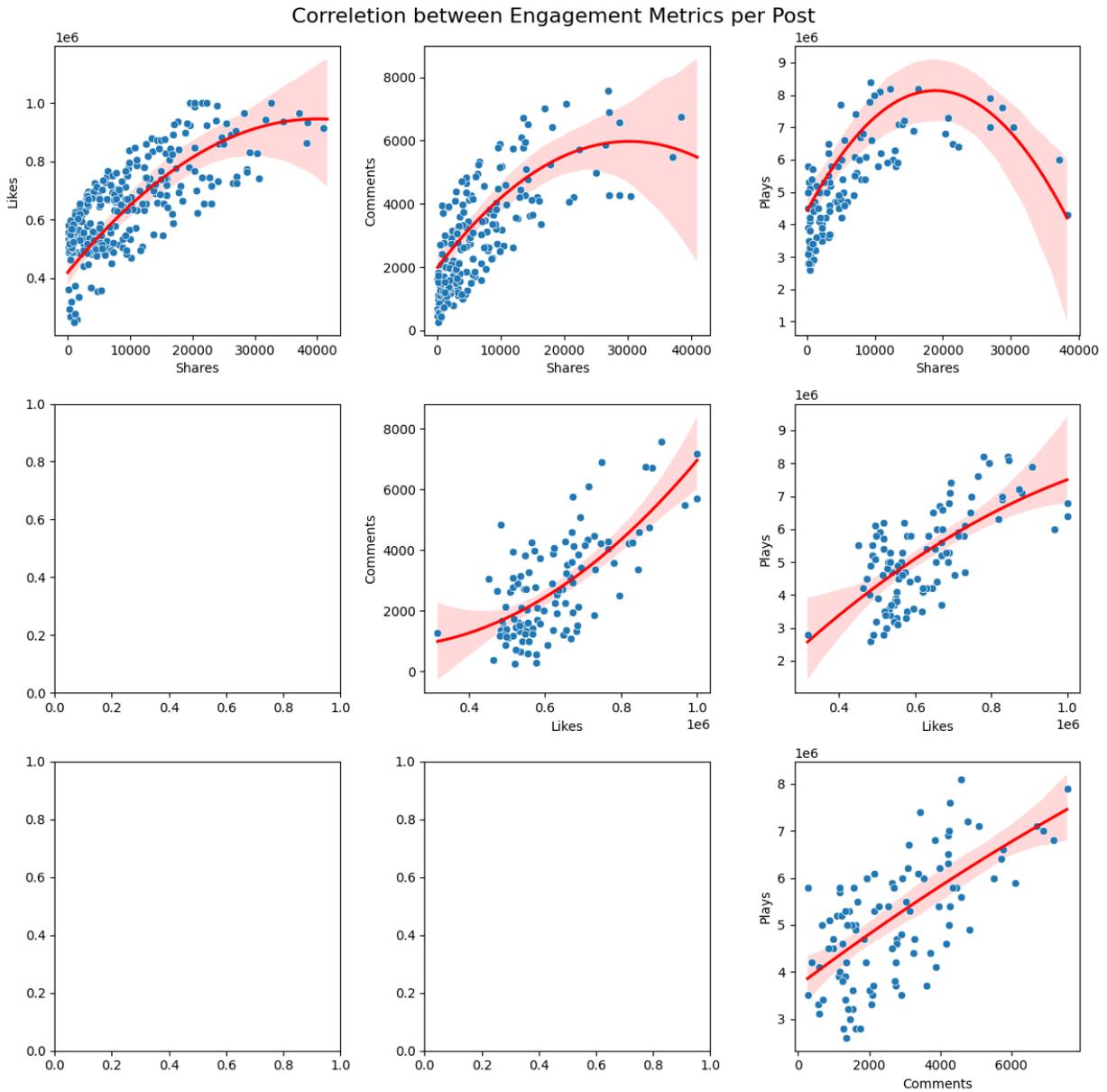

**Fig 1.** *In this graph, I cross-correlated each of the 4 virality metrics (shares, likes, plays, number of comments) for each post for the 2000 posts that I scraped.*



For this study, I focused on 4 main virality metrics which consisted of share count, play count, number of likes, and number of comments for each post. I correlated each one of these key metrics with each other to shed light on the dynamics of content virality on TikTok. I found that the number of comments, likes, and shares exhibited a positive correlation with each other. This suggests that posts that perform well in one aspect tend to perform well in others, indicating a holistic relationship between engagement factors. Intuitively, this makes sense, as posts that gain traction with a target audience are more likely to garner a response from them in the comments section.

However, correlation analysis seems to uncover a reciprocal relationship between sharing and interactions after a certain point. Up to a certain point, as the number of shares on a post increased, so did the overall interactions and views. This suggests that sharing plays a pivotal role in amplifying the reach and engagement of a TikTok post within the tech product niche. A higher view count was also associated with more sharing. This implies that the visibility and attractiveness of a video, perhaps due to its engaging content or appeal, can encourage viewers to share it with their networks, creating a positive feedback loop that contributes to its virality (Quesenberry). However, after a certain midway point of the number of shares for a post, the overall interactions seem to teeter off. This suggests that after a while, an incredibly viral post probably has been viewed so many times that it starts getting reshared to the same audience, resulting in diminishing returns in terms of new interactions and views. This insight underscores the importance of not only achieving virality but also sustaining it, as overexposure to the same audience can eventually lead to a saturation point in a post's performance (Quesenberry).



Community Activity

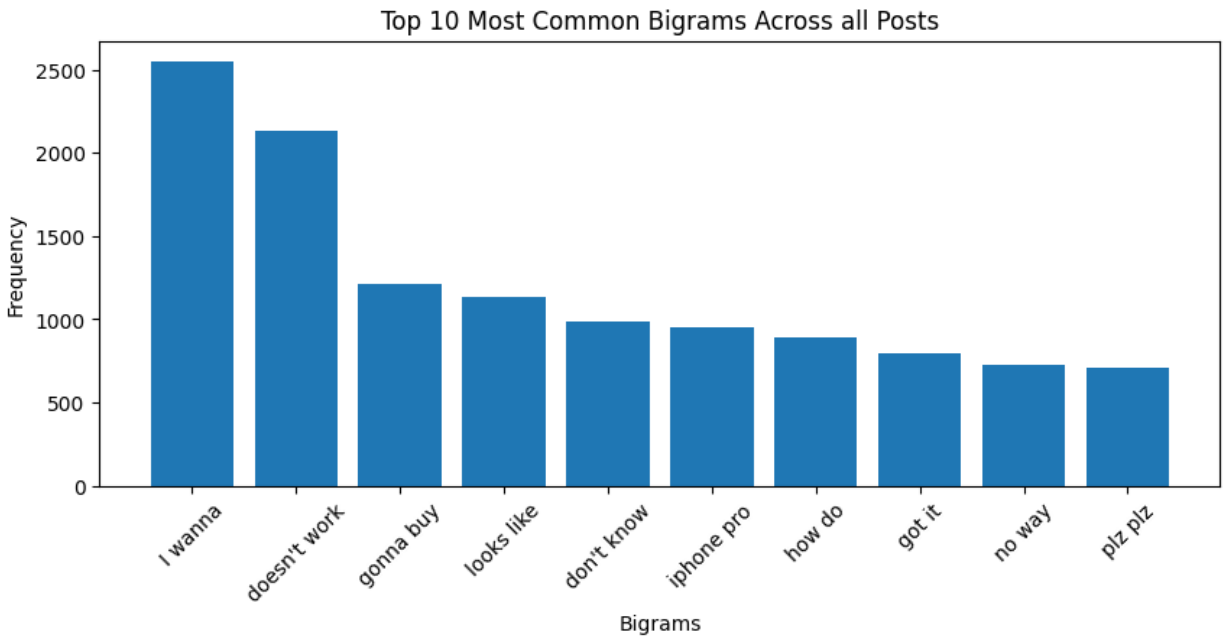

**Fig 2.** *The figure above shows the frequency in units of 100 of the most common bigrams that were collected from all 274,651 comments.*

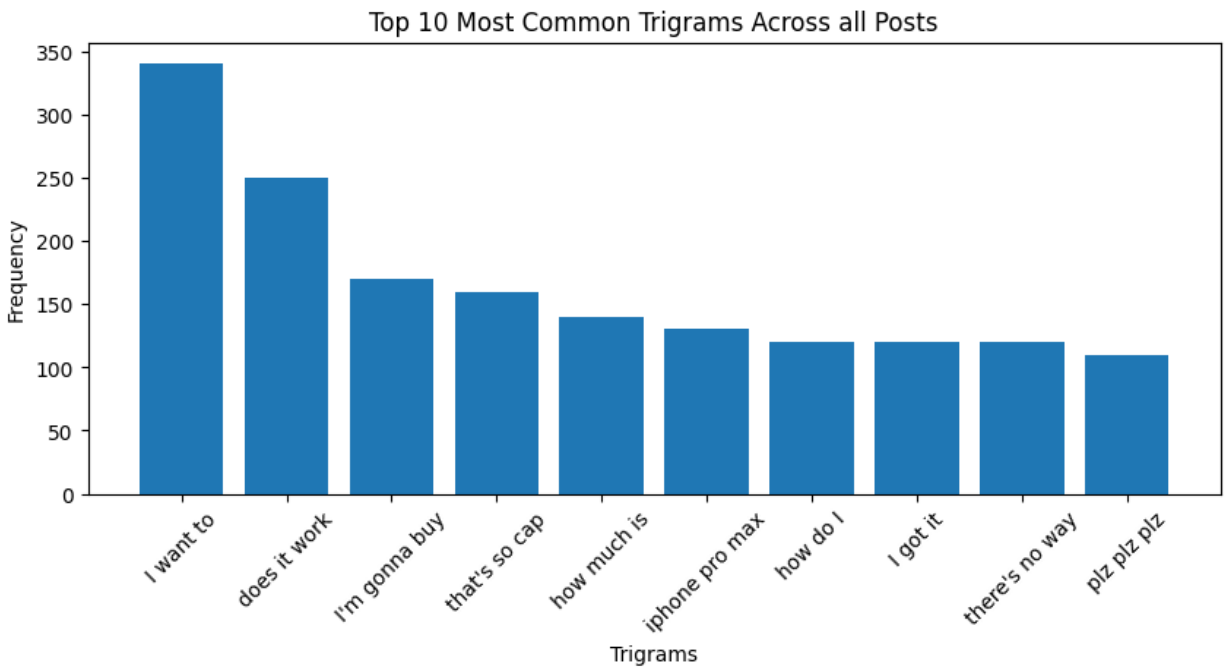

**Fig 3.** *This graph above shows the frequency in units of 100 of the most common trigrams that were collected from all 274,651 comments.*



Bigrams and trigrams play a crucial role in sentiment analysis when evaluating comments because they provide a more nuanced understanding of the language used by users. Analyzing individual words in isolation can often lead to misinterpretations of sentiment, as the meaning of a word can change dramatically when combined with other words. Bigrams, which are pairs of consecutive words, and trigrams, which are sequences of three consecutive words, capture these contextual nuances. For instance, the word "not" followed by a negative word like "bad" would indicate a positive sentiment, whereas "not" followed by a positive word like "good" would reinforce a negative sentiment. By considering these multi-word combinations, sentiment analysis algorithms can more accurately discern the sentiment of a comment, taking into account the subtleties of language that would be missed if only single words were analyzed. This leads to more accurate and reliable sentiment classification, which is invaluable for understanding user opinions and feedback in various domains, from product reviews to social media discussions. I collected bigrams and trigrams by removing all stop words, as this preprocessing step helps focus the analysis on meaningful word combinations. Stop words, such as "and" "the" "in," and "of" are commonly occurring words that do not carry specific sentiment or meaning on their own. I used a hash table to extract combinations of two and three word pairings that were the most most commonly consecutively occurring.

Several intriguing patterns and insights emerged from analyzing bigrams and trigrams., shedding light on the dynamics of user engagement and sentiment within the tech product niche allowing this research to extract information on eWOW. One of the prominent n-grams, "I wanna" or "I want to", reflects a common purchase intention sentiment prevalent in these videos. It's noteworthy that the majority of the videos in the dataset either featured technology devices or



services, aligning with users expressing their desire to obtain or experience the showcased

products or services. This indicates a correlation between the content and users' purchase intent,

a vital aspect of viral posts. This suggests that users on social media platforms are not only

expressing a desire for products but are also actively seeking information about the latest and

best in technology. The inclination towards newer and trending technology reflects the dynamic

nature of user interests and the importance of staying current in content creation for higher

engagement and virality (Mou).

The gram "doesn't work" serves as a revealing indicator of the skepticism deeply

ingrained among TikTok viewers in the tech product niche. This skepticism, while multifaceted,

can be categorized into three primary dimensions (Chaudhary). First, it often stems from genuine

disbelief in the efficacy of the advertised product or service. Users expressing this sentiment may

have reservations about the product's claims or may have encountered negative reviews or

experiences with similar offerings in the past. Second, the prevalence of comments containing

"doesn't work" can be attributed to individuals who have no intention of trying the technology

product themselves and may engage in discouraging others from making a purchase. This

behavior, while not necessarily malicious, underscores the influence that negative comments can

have on potential consumers. With more data on comments, it would be interesting to explore if

more successful content creators consider strategies to counter such narratives, potentially by

showcasing satisfied customers or addressing common concerns. The third dimension of

skepticism related to the "doesn't work" bigram is the apprehension surrounding the reliability

and quality of the product or service being advertised. With technology products, each new

device usually offers new state-of-the-art features which might induce skepticism on whether or



not the product actually works as advertised. With social media in general, it's really easy for creators to use video-editing and post-processing to falsely advertise a product's functionality. When viewers voice uncertainties regarding a product's functionality or reliability, it frequently initiates extensive comment threads where users exchange their personal experiences, encompassing both positive and negative encounters. These interactions provide an invaluable platform for viewers to thoroughly evaluate a tech product, gather supplementary insights, and potentially address any concerns. Viewers are more likely to trust people like themselves who share similar concerns or questions regarding the functionality of a phone (Zhang). In this way, these discussions can transform initial skepticism into a foundation of trust and foster lasting loyalty among the audience.

The gram "gonna buy", or "i'm gonna buy", highlights the intriguing concept of purchase intention in two forms: those expressing intent to purchase and those expressing intent not to. What's particularly fascinating is the observation that these comments often follow a pattern of mimicking one another. This phenomenon can be attributed to a form of mob mentality, where early comments create a trend that subsequently influences others to adopt a similar stance (Krumm). Understanding this dynamic is crucial because it means the initial reaction to a post is seminal for how the rest of communication reacts and interacts with the content. More people are likely to buy a phone if people are raving about it in the comments.

Lastly, the bigram "looks like" introduces a novel concept of similarity and comparison in evaluating the promoted products or services. Many comments allude to the idea that viewers tend to compare the showcased item to something they already know. This implies that viewers use existing reference points to assess the value proposition of the product. This bigram was



more prevalent in videos that showcased some unreleased Chinese Android phones. The comments that evaluate the uniqueness of the phone seem to repeatedly compare the device with the phones that have been well-established in the market and are usually regarded as the gold standard for flagship smartphones. A discernible sentiment of dissatisfaction becomes apparent when these smartphones fall short of matching the extensive features typically associated with flagship products offered by industry giants like Samsung and Apple. Consumers often expect a certain level of performance, innovation, and premium features from these established brands, which can lead to disappointment when alternative devices do not meet those high standards. However, it's equally noteworthy that users exhibit a sense of gratification when these newer, possibly lesser-known devices are showcased to bring something fresh and unique to the table. Whether it's through groundbreaking technology, novel design elements, or unexpected functionalities, these offerings can pique the interest and curiosity of tech enthusiasts. In such instances, viewers are not only open to embracing the innovation but may also view these devices as potential challengers to the established market leaders. This dichotomy in viewer sentiment underscores the competitive landscape of the tech product niche on platforms like TikTok. It highlights the importance of not only meeting but exceeding user expectations to garner a positive response, all while keeping an eye on the ever-evolving trends and preferences that can elevate a product to newfound prominence in the highly competitive world of tech innovation.

| Xiaomi 13t | Huawei P60 Pro | OPPO Find X5 Pro |
|---|---|---|
| "Camera looks like (Samsung) S23" | "No Bezel looks soooo much better like new iphone" | "Spider eye camera… what's new, same as 3 year old iPhone…" |
| "… almost looks same like | "Hope metallic finish is a | "Too much editing… bad |



| other Samsung phones" | little less slippery, looks scary like iPhone dilemna" | photos like computational photography gimmick on Samsung" |
|---|---|---|

**Table 2.** *This table shows a snippet of some of the comments that were pulled from TikTok posts regarding Chinese-Manufactured Android phones. The international models of these devices are often purchased through third-party retailers and used as unlocked cellular devices in the United States (Xu). The frame of comparison for these phones are mostly the flagship phones manufactured by Apple and Samsung, which are the 2 leading smartphone manufacturers in the US.*

Certain bigrams and trigrams in TikTok comments often reiterate common nonsensical expressions, reflecting a recurring pattern in user engagement dynamics. This phenomenon becomes particularly evident as a post gains traction and starts to go viral. At this point, people are not only drawn to engage with the content but also motivated by the desire to affiliate themselves with similar videos and themes prevalent in the TikTok ecosystem (Zhu).

When individuals encounter a post they resonate with or one that aligns with their preferences, be it a tech product showcase or any other niche, they are more inclined to take action beyond passive viewing. They save the post, drop comments, or share it with their followers. This behavior serves a dual purpose. Firstly, it functions as a bookmarking mechanism, allowing users to revisit the content later or easily access it for future reference. Secondly, and perhaps more intriguingly, it's a means of signaling their interest in similar content to both the TikTok algorithm and their own followers. By engaging with a post, users effectively communicate their desire for more content of a similar nature. The TikTok algorithm, which thrives on user engagement data, takes note of these interactions and subsequently tailors the user's feed to include related videos (Zhang). With tech products, some of the special features have very technical names that are not necessarily known to viewers. Not everyone knows what a Pro Motion 120Hz display or USB-C Fast Charging with handshakes necessarily is off the top



of their heads. If the content piques their interest, they are more reliant on the recommendation algorithm to share similar content instead of manually searching it up themselves. Simultaneously, when users comment on or share a post, it circulates the content among their network, fostering a sense of community around a specific theme or trend.

For performing sentiment analysis over the comments, I initially opted in to using the LLM trained off Reddit data to identify 28 different emotions. However, the comments were too sparse in order to draw proper trends. I opted in using the VADER instead which instead gave me a continuous value expressing the overall sentiment of the phrase. The value was either positive, zero, or negative which corresponds to a range of emotion as expressed in the table below. I ran the VADER model on every single comment that I scraped. I then averaged the sentiment score for each post and plotted it below to understand how sentiment affects the virality metric.

|  | Negative | Zero | Positive |
|---|---|---|---|
| Emotions | Sorrow, embarrassment, anger, fear | Neutral, curiosity, desire | Happy, excitement, amusement, admiration, pride, optimism |

**Table 3.** *Using the continuous value from the VADER model essentially collapses the 28 emotion, down to 3 which allows for a bit more expressive modeling when it comes to the roles of sentiment and the virality of tech products on Tiktok.*



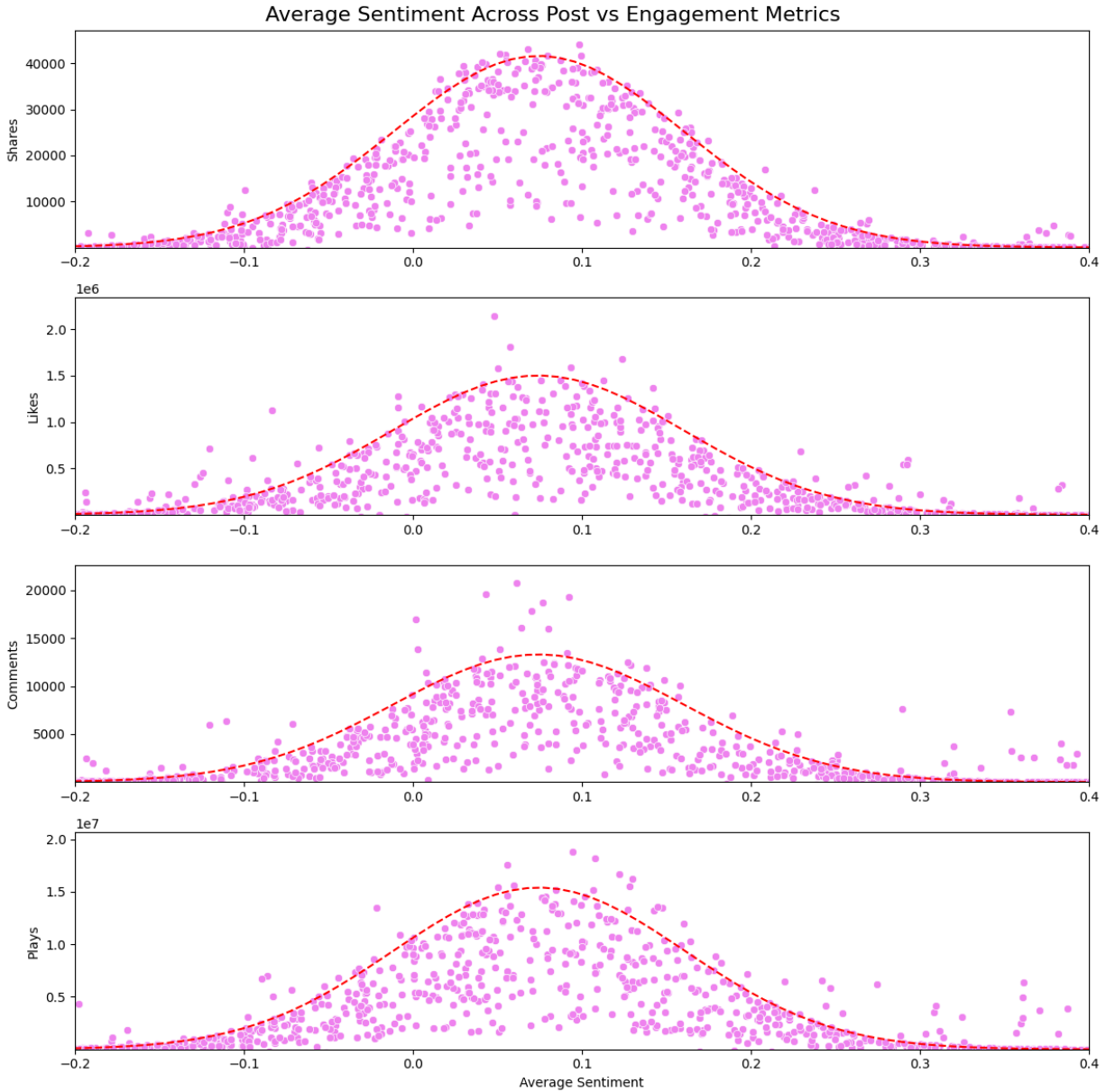

**Fig 4.** *I graphed each engagement metric with respect to the average sentiment score for every post.*

The graphical analysis of virality metrics in relation to the average sentiment per post reveals an intriguing pattern: posts with a moderately positive sentiment tend to exhibit the highest levels of virality. It's important to note that this sentiment is not entirely neutral; there's a discernible lean towards a mildly positive sentiment in the comments. This trend aligns with the



intuitive understanding that posts accompanied by overly negative comments are often overlooked, as they may contain harmful or contentious content, which TikTok's community tends to avoid. Similarly, posts accompanied by overwhelmingly positive comments can also face a potential downside, as they may be perceived as spammy or insincere by viewers.

The emphasis on a slightly positive sentiment in driving virality suggests that TikTok users are more receptive to tech products that strikes a balance between positivity and authenticity. Authenticity in the context of social media posts and influencers refers to the degree of genuineness, transparency, and honesty in the content shared online. It involves the alignment between a person's true personality, values, and beliefs, and the content they create and share with their audience, fostering trust and credibility. Such content likely resonates with a broader audience, offering engagement without veering into extremes of emotion.  This aligns with TikTok's ethos of fostering an engaging and enjoyable user experience while avoiding divisive or overly promotional content (Mansoor). Having a controversial product that garners equal parts negative and positive sentiment might also be more suitable for the recommendation algorithm, as it is more likely to recommend such content to users who might engage similarly.



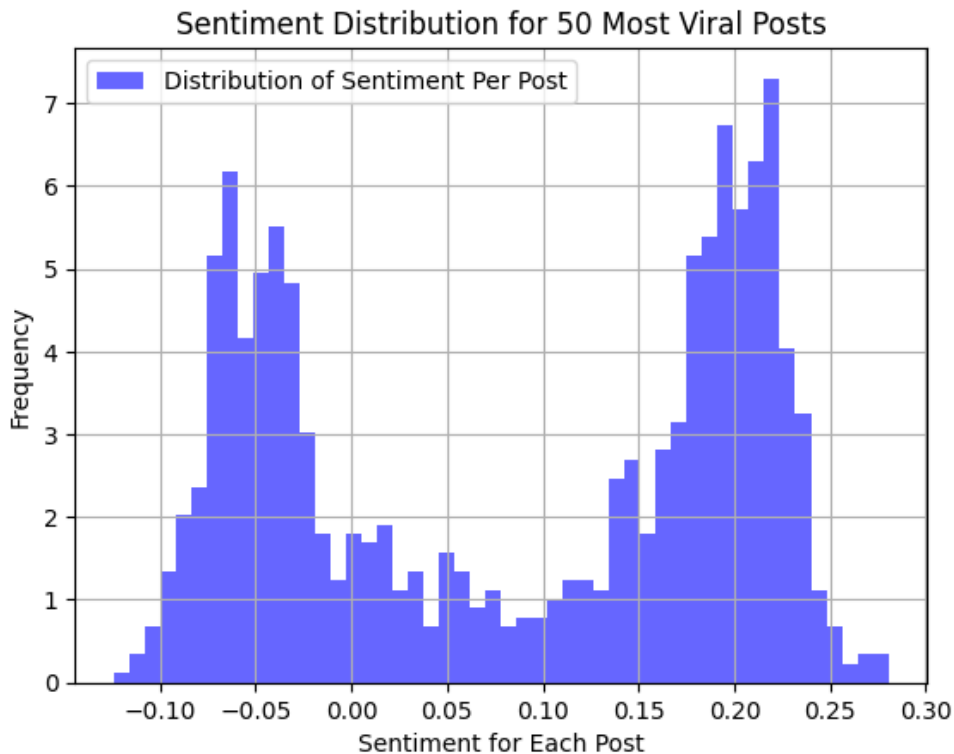

**Fig 5.** *Plotting the sentiment for every content for the 50 most viral posts from our dataset shows a very clear bimodal distribution between comments that are perceived to be negative and those that are negative.*

When I delved into the analysis of the top 50 viral posts, a fascinating pattern emerges: the comments within a single post exhibit a bimodal distribution in nature. This means that, on average, viral posts tend to attract an equal amount of both negative and positive comments. This phenomenon aligns with the inherent nature of viral content, which often aims to elicit strong emotional reactions from viewers. When content is either entirely positive or entirely negative, it may fail to engage audiences effectively because it lacks the diverse and dynamic qualities that stimulate interaction.



The idea behind this bimodality is rooted in the understanding that viral content thrives on controversy and intrigue. When a tech product depicted in a post generates a polarizing response, it tends to spark curiosity and ignite discussions among viewers. This controversy not only amplifies engagement but also broadens the spectrum of opinions and perspectives surrounding the content. People are more likely to go out of their way and try to learn more about this product if it has controversy. Such diverse discussions draw more attention to the post, effectively increasing its virality.



Influencer

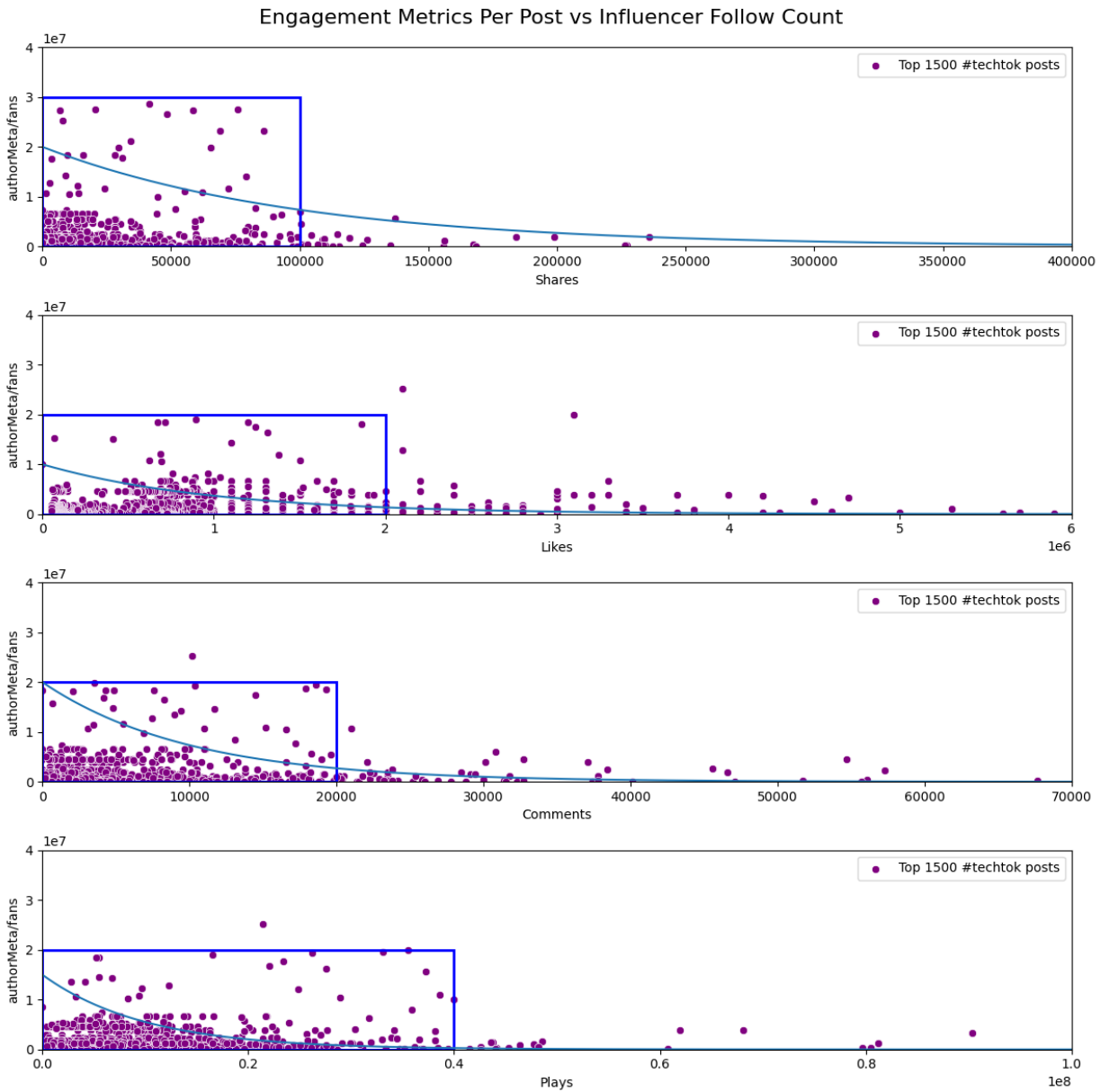

**Fig 6.** *I take the top 1500 posts about tech products from our dataset and correlate it with influencer statistics.*

For analyzing influencer trends, I took the top 1500 posts in our dataset and correlated influencer metadata with the engagement metrics. For the sake of this research, I considered any creator with a history of posting content that spans more than one occasion. The goal is see if



there are specific factors from an influencer's profile that are deterministic in the virality of their contents. The main metric for influencers that I can see is their total number of likes and number of followers.

It becomes quite evident that posts garnering exceptionally high popularity metrics often did not have any specific association with creators who boasted a substantial following. In fact, the posts that achieved superlative virality seemed to emerge rather spontaneously, lacking any inherent connection to users who consistently produce viral content or possess a track record of accumulating a massive follower base. What's particularly intriguing is the observation that creators with substantial followings typically exhibited merely average levels of viewership, engagement, and resharing from their established audience. While this observation hints at a possible correlation, it doesn't firmly establish the notion that a creator's past history of virality necessarily enhances their ability to generate content deemed viral. This observation challenges the conventional belief that a creator's previous success in producing viral content guarantees a repeat performance. Instead, it suggests that viral content creation is a multifaceted endeavor that cannot be solely attributed to a creator's follower count or prior achievements. Virality hinges on various factors, including content novelty, timing, audience resonance, and sheer luck. Therefore, even creators with significant followings must continually innovate and tap into current trends to increase their chances of producing content that captures the viral essence of the moment.

Exceptional virality does not necessarily align with the typical influencer profile. Influencers who achieve extreme virality often adopt a consistent upload schedule only after sustaining their initial success. Upon implementing a more regular content release strategy, there is a noticeable trend of increased engagement and viewership. However, this appears to be



indicative of organic audience growth. When content resonates strongly with viewers and aligns with the recommendations of the algorithm, it leads to a gradual expansion of viewership. The algorithm continues to display the content to profiles similar to the existing audience, creating a self-reinforcing cycle.

Influencers who experience extremely viral videos tend to have more polished accounts. If an influencer consistently creates content that attracts substantial viewership and engagement, it suggests that content creation likely constitutes their full-time occupation. This enables them to invest time and resources in their content creation process. They may engage in research before creating content on a specific topic and dedicate effort to producing refined TikTok videos. This may involve setting up the camera, providing narration, or crafting a compelling narrative throughout the post. There are only so many different ways that creators can augment and portray a computer or a smart home accessory. But having a consistent background set and overall polished presentation approach are just some of the way viewers might identify a tech blogging channel to be more refined and professional.

The growth in followership is a significant outcome of creating viral content. People are more likely to follow a creator whose content they genuinely enjoy, as they have confidence that the creator will continue producing similar content in the future. Therefore, followers represent the long-term growth trajectory resulting from viral content. This increased followership is a testament to the creator's ability to capture and retain the interest of their audience, solidifying their presence and influence on the platform.



<u>Content</u>

Despite the notable strides made in multi-modal large language models, our approach involved a manual examination of the TikTok posts' content. I chose this method to ensure that our analysis would encompass a richer and more nuanced perspective, primarily because there lacked a well-defined framework for codifying and categorizing the diverse content available. Our overarching objective when scrutinizing the videos and slideshows was to uncover the commonalities and distinctions among various visual formats embedded within the dataset of 1,500 posts. This approach allowed us to delve deeply into the content, transcending the limitations of automated methods by capturing subtleties and contextual nuances that might have been missed. I aimed to provide a more comprehensive understanding of the TikTok posts, taking into account the intricacies of visual storytelling, creative techniques, and audience engagement strategies employed by content creators. I sought to unearth patterns, trends, and insights that might not have been discernible through automated processes alone, contributing to a more holistic and expressive examination of the diverse tech content landscape.

|  | Corporate News | Novel Features | Aesthetic Tech Products | How To/DIY Guides |
|---|---|---|---|---|
| **Significance** | Sought to inform the public about news that might affect technology products or services | Aims to showcase a product that can provide a new function to its end-users | Products that are merely collectibles, have aesthetic qualities | Short-form videos that aim to show viewers how they can achieve life-hacks |
| **Example** | Tesla Model 3 refresh, Apple stock stagnating | iPhone 14 Pro 48 Megapixel camera, Foldable | RGB Smart Home Lights, Automatic Cat Litter Box | How to build a standing desk, how to make a |



| | after earning call | Samsung phone | | tiny robot arm |
|---|---|---|---|---|

**Table 4:** *Manually codifying the 2000 posts resulted in 4 main themes. Each theme is unique from each other and the content in each is meant to evoke a different kind of reaction in the audience.*

The first category of posts feature either innovative technologies or novel applications of existing technology that were unfamiliar to the audience. When content showcases these inventive elements, it tends to spark viewers' interest and intrigue, making them more inclined to share it with their networks. The key driver behind this behavior is the perception of ingenuity and novelty. When viewers come across content they perceive as ingenious and groundbreaking, they are not only eager to share it but also more open to the idea of incorporating the featured technology into their own lives. This phenomenon underscores the significant role of innovation in driving content virality. Audiences are drawn to fresh ideas and inventive solutions that address unmet needs or fill gaps in their daily lives. As a result, content creators who introduce unique technological concepts or innovative uses of technology stand a higher chance of not only gaining traction but also fostering audience engagement and potential adoption of the showcased technology. This highlights the importance of creativity and novelty as potent factors in the virality and influence of tech-related content on platforms like TikTok.

Our second category of posts feature exclusive or breaking news snippets, positioning the creators as pioneers in disseminating this information on TikTok. This pioneering role as early news sharers translated into them being the primary sources of this news within the platform. Consequently, viewers who came across these posts were more inclined to share them with their own circles, establishing these posts as influential nodes in the realm of electronic word-of-mouth. This phenomenon reflects the vital role of timely and exclusive content in



catalyzing virality. When creators are among the first to share noteworthy news or information, their posts gain an inherent edge in attracting attention and engagement. Viewers recognize the value of being among the first to disseminate such content, and this prompts them to share it within their networks, triggering a cascading effect that fuels virality. Especially with tech products, there is constant innovation and research being performed in the industry. Therefore, in the context of viral content dynamics, being a pioneer in sharing breaking news can elevate a post's status, turning it into a focal point of electronic word-of-mouth conversations.

In our third category, the content revolves around products that exhibit exceptional aesthetic appeal. These items are characterized by their visually striking design, often exuding an air of luxury and exclusivity. They possess a certain allure that captivates viewers, creating a desire to possess them, even if they come with a hefty price tag that is beyond the reach of most individuals. In such instances, these products serve as aspirational objects, prompting viewers to admire them and yearn for their own possession. Conversely, another category of content features products or services that serve as budget-friendly alternatives to their more expensive counterparts, commonly referred to as "dupes." These offerings provide a practical and affordable solution for consumers who seek the same benefits or aesthetics as premium products without breaking the bank. These dupes cater to a more budget-conscious audience, resonating with those who appreciate value and practicality. The dynamic between content that showcases aesthetically pleasing, high-end products and content highlighting budget-friendly alternatives is intriguing. While the former fuels desires for luxury and exclusivity, the latter taps into the practicality and accessibility that resonate with a broader audience. Both types of content serve as influential drivers within the realm of consumer preferences and choices.



Our last category of content falls into the category of DIY (Do It Yourself) guides, presenting viewers with concise and instructional content that guides them through specific tasks or projects. These DIY videos aim to empower viewers by offering step-by-step instructions on how to accomplish various activities, from crafting and cooking to home improvement and life hacks. What sets the popular DIY videos apart is their ability to introduce ingenious life hacks that require minimal materials and time to set up. The allure of these DIY videos lies in their capacity to provide practical solutions to everyday challenges in a simple and innovative manner. Viewers are drawn to content that presents them with clever and time-saving shortcuts or creative problem-solving approaches. The popularity of such videos stems from their ability to resonate with a wide audience seeking quick and efficient solutions to common issues.

**CONCLUSION**

<u>Summary of Findings</u>

In conclusion, this thesis reveals that the virality of tech products on TikTok is a complex, multi-dimensional phenomenon. It underscores the importance of engagement metrics like comments, likes, and shares in driving virality. The data shows an inclination towards sharing playing a pivotal role in amplifying the reach and engagement of a TikTok post within the tech product niche. The analysis of user comments, through bigrams and trigrams, provides insights into consumer sentiments, especially purchase intentions. More specifically, there is an increasing prevalence in purchase intention for a majority of the posts that went viral. There is also an occurrence of heavy skepticism within community activity, which shows that tech products that draw controversy are more likely to spread in different social circles. This study



also highlights the nuanced role of influencers, where their impact on content virality is not solely dependent on follower counts, but also on factors like content novelty and audience engagement. This suggests that achieving virality for tech products on TikTok requires a strategic approach, balancing content creativity, influencer collaboration, and a deep understanding of audience dynamics.

Further Study

Near the completion of this thesis, Tiktok released a storefront as a novel feature, The TikTok store. This is a feature on the TikTok platform that allows users to shop directly within the app. This is part of the broader trend of social commerce, where social media platforms integrate e-commerce capabilities. Users can browse and purchase products without leaving the TikTok app, making the shopping experience more seamless and integrated with the content they enjoy (OpenAi). This feature caters to the convenience of users and taps into the influencer marketing model, as creators can promote products directly to their followers. This could be a powerful tool for future studies that aim to understand purchase intention through community activity.

If this study were to be repeated with, it would be invaluable to construct a dataset that was more substantial in size. Due to budget and time constraints imposed on the scope of the project, I was limited to scraping about 100-200 comments per TikTok post. Although sufficient for overarching claims, it would be fruitful to perform more robust time series analysis over a larger dataset. Scraping all the comments for a singular post might uncover nuances and subtleties for sentiments on community activity. In a time series analysis, I can unravel some of the more enduring patterns regarding the initial set of comments. It would be valuable to explore



if comments posted in the early stages of the post's existence are more likely to be repeated or

not.




**REFERENCES**

Bakhshi, S., Shamma, D. A., & Gilbert, E. (2014, April). Faces engage us: Photos with faces attract more likes and comments on instagram. In Proceedings of the SIGCHI conference on human factors in computing systems (pp. 965-974).

Berger, J., & Milkman, K. L. (2012). What makes online content viral?. Journal of marketing research, 49(2), 192-205.

Berger, J. (2016). Contagious: Why things catch on. Simon and Schuster.

Botha, E., & Reyneke, M. (2013). To share or not to share: the role of content and emotion in viral marketing. Journal of Public Affairs, 13(2), 160-171.

Choudhary, Nilam, Chitra Gautam, and Vivek Arya. "Digital marketing challenge and opportunity with reference to tiktok-a new rising social media platform." Editorial Board 9.10 (2020): 189-197.

Chen, Roger, and Eric PH Li. Growing and Shaping the User Base: The Transformation of TikTok. SAGE Publications: SAGE Business Cases Originals, 2023.

Chevalier, Judith A., and Dina Mayzlin. (2006). The effect of word of mouth on sales: Online book reviews. Journal of Marketing Research 43.3: 345–354.

de la Roche, Caitlin, et al. "An investigation on consumer perceptions of email and social media marketing: An advertising case in South Africa." International Review of Management and Marketing 12.4 (2022): 29.

De Vries, L., Gensler, S., & Leeflang, P. S. (2012). Popularity of brand posts on brand fan pages: An investigation of the effects of social media marketing. Journal of Interactive Marketing, 26(2), 83-91.





Djafarova, E., & Rushworth, C. (2017). Exploring the credibility of online celebrities' Instagram profiles in influencing the purchase decisions of young female users. Computers in Human Behavior, 68, 1-7.

Elmas, T., Stephane, S., & Houssiaux, C. (2023, April). Measuring and Detecting Virality on Social Media: The Case of Twitter's Viral Tweets Topic. In Companion Proceedings of the ACM Web Conference 2023 (pp. 314-317).

Goel, S., Anderson, A., Hofman, J., & Watts, D. J. (2016). The structural virality of online diffusion. Management Science, 62(1), 180-196.

Grure, L., & Rasmussen, V. (2023, July 5). *#Beautytok going viral*. AURA. Retrieved September 13, 2023, from https://uia.brage.unit.no/uia-xmlui/handle/11250/3082365

Jenkins, H., Ford, S., & Green, J. (2013). Spreadable media. In Spreadable media. New York University Press.

Kaplan, A., & Haenlein, M. (2011). Two hearts in three-quarter time: How to waltz the social media/viral marketing dance. *Business Horizons*, *54*(3), 253-263.

https://www.sciencedirect.com/science/article/abs/pii/S0007681311000152

Kulkarni, K. K., Kalro, A. D., & Sharma, D. (2020). The interaction effect of ad appeal and need for cognition on consumers' intentions to share viral advertisements. Journal of Consumer Behaviour, 19(4), 327-338.

Krumm, Justine S. "Influence of social media on crowd behavior and the operational environment." ARMY COMMAND AND GENERAL STAFF COLLEGE FORT LEAVENWORTH KS SCHOOL OF..., Tech. Rep (2013).




Li, F., Larimo, J. & Leonidou, L.C. Social media marketing strategy: definition, conceptualization, taxonomy, validation, and future agenda. J. of the Acad. Mark. Sci. 49, 51–70 (2021). https://doi.org/10.1007/s11747-020-00733-3

Ling, C., Blackburn, J., De Cristofaro, E., & Stringhini, G. (2022, June). Slapping cats, bopping heads, and oreo shakes: Understanding indicators of virality in tiktok short videos. In Proceedings of the 14th ACM Web Science Conference 2022 (pp. 164-173).

Liu, Z., Zou, L., Zou, X., Wang, C., Zhang, B., Tang, D., ... & Cheng, Y. (2022). Monolith: real time recommendation system with collisionless embedding table. arXiv preprint arXiv:2209.07663.

Lou, C., & Yuan, S. (2019). Influencer marketing: How message value and credibility affect consumer trust of branded content on social media. Journal of Interactive Advertising, 19(1), 58-73.

Mansoor, Iqbal (Apr. 2023). Tiktok revenue and Usage Statistics (2023). url: https://www.businessofapps.com/data/tik-tok-statistics/.6

Minton, E., Lee, C., Orth, U., Kim, C.-H., & Kahle, L. (2012). SUSTAINABLE MARKETING AND SOCIAL MEDIA: A Cross-Country Analysis of Motives for Sustainable Behaviors. *Journal of Advertising*, *41*(4), 69–84. http://www.jstor.org/stable/23410034

Mou, Jessie Boxin. Study on social media marketing campaign strategy--TikTok and Instagram. Diss. Massachusetts Institute of Technology, 2020.

Nelson-Field, K., Riebe, E., & Newstead, K. (2013). The emotions that drive viral video. Australasian Marketing Journal (AMJ), 21(4), 205-211.



Permatasari, A., & Kuswadi, E. (2017). The impact of social media on consumers' purchase
intention: A study of ecommerce sites in Jakarta, Indonesia. Review of Integrative
Business and Economics Research, 6, 321.

Petrescu, M., Korgaonkar, P., & Gironda, J. (2015). Viral advertising: A field experiment on viral
intentions and purchase intentions. Journal of Internet Commerce, 14(3), 384-405.

Petty, R. E., & Cacioppo, J. T. (1986). The Elaboration Likelihood Model of persuasion. In L.
Berkowitz (Ed.), Advances in experimental social psychology (Vol. 19, pp. 123-205).
Academic Press.

Phelps, J. E., Lewis, R., Mobilio, L., Perry, D., & Raman, N. (2004). Viral marketing or
electronic word-of-mouth advertising: Examining consumer responses and motivations to
pass along email. Journal of advertising research, 44(4), 333-348.

Quesenberry, Keith A., and Michael K. Coolsen. "What makes Facebook brand posts engaging?
A content analysis of Facebook brand post text that increases shares, likes, and comments
to influence organic viral reach." Journal of Current Issues & Research in Advertising
40.3 (2019): 229-244.

Segev, S., & Fernandes, J. (2023). The anatomy of viral advertising: A content analysis of viral
advertising from the elaboration likelihood model perspective. Journal of Promotion
Management, 29(1), 125-154.

Stephen, A. T. (2016). The role of digital and social media marketing in consumer behavior.
Current opinión in Psychology, 10, 17-21.

Worb, J. (2023, February 16). How Does The TikTok Algorithm Work? Here's What You Need
To Know.




Https://Later.com/Blog/Tiktok-Algorithm/#:~:Text=According%20to%20TikTok%3A%2 0%E2%80%9CThe%20system,'Re%20not%20interested%20in.%E2%80%9D.

https://later.com/blog/tiktok-algorithm/#:~:text=According%20to%20TikTok%3A%20% E2%80%9CThe%20system

Xu, Fangqi. "A Smartphone Challenger's Competitive Strategy: The Case of Xiaomi." Kindai management review 3.1 (2015): 90-99.

Zarrella, D. (2009). The social media marketing book. " O'Reilly Media, Inc.".

Zhang, Wenyao, Wei Zhang, and Tugrul U. Daim. "Investigating consumer purchase intention in online social media marketing: A case study of Tiktok." Technology in Society 74 (2023): 102289.

Zhang, Min, and Yiqun Liu. "A commentary of TikTok recommendation algorithms in MIT Technology Review 2021." Fundamental Research 1.6 (2021): 846-847.

Zhu, Yu-Qian, and Houn-Gee Chen. "Social media and human need satisfaction: Implications for social media marketing." Business horizons 58.3 (2015): 335-345.